\documentclass[prl,twocolumn,superscriptaddress,showpacs]{revtex4}
\usepackage{amsmath}
\usepackage{amsfonts}

\begin{document}
\title{Classical and quantum communication without a shared reference
  frame}
\author{Stephen D. Bartlett}
\email{bartlett@ics.mq.edu.au}
\affiliation{Department of Physics, Macquarie University, Sydney, New
  South Wales 2109, Australia}
\author{Terry Rudolph}
\email{rudolpht@bell-labs.com}
\affiliation{Bell Labs, 600-700 Mountain Ave., Murray Hill, NJ 07974,
  U.S.A.}
\author{Robert W. Spekkens}
\email{rspekkens@perimeterinstitute.ca}
\affiliation{Perimeter Institute for Theoretical Physics, Waterloo,
  Ontario N2J 2W9, Canada}
\date{7 July 2003}

\begin{abstract}
  We show that communication without a shared reference frame is
  possible using entangled states.  Both classical and quantum
  information can be communicated with perfect fidelity without a
  shared reference frame at a rate that asymptotically approaches one
  classical bit or one encoded qubit per transmitted qubit.  We
  present an optical scheme to communicate classical bits without a
  shared reference frame using entangled photon pairs and linear
  optical Bell state measurements.
\end{abstract}
\maketitle

Quantum physics allows for powerful new communication tasks that are
not possible classically, such as secure communication~\cite{BB84} and
entanglement-enhanced classical communication~\cite{Mac02}.  In
investigations of these and other communication tasks, considerable
effort has been devoted to identifying the \emph{physical} resources
that are required for their implementation. It is generally presumed,
at least implicitly, that a shared reference frame (SRF) between the
communicating parties is such a resource, with the precise nature of
the reference frame being dictated by the particular physical systems
involved. For example, if the sender (Alice) and receiver (Bob) are
communicating via spin-1/2 systems, it is generally presumed that they
must share a reference frame for spatial orientation so that they may
prepare and measure spin components relative to this frame. Despite
the ubiquity of this presumption, we shall be asking whether it is
\emph{necessitated} by the laws of quantum physics.

This question is clearly of interest for pragmatic reasons.
Establishing a SRF between two parties requires communication via a
channel that is capable of transmitting some ``physical'' information
(such as spatial orientation). Establishing a \emph{perfect} SRF
requires infinite communication (i.e., transmitting a system with an
infinite-dimensional Hilbert space, or an infinite number of systems
with finite-dimensional Hilbert spaces~\cite{Per01,Per01b}.) Moreover,
any finite (i.e., imperfect) SRF must be treated quantum mechanically
and thus inevitably suffers disturbances during measurements, causing
it to degrade.  Finally, we note that shared prior entanglement, a
valuable resource in distributed quantum information processing, can
be consumed to establish SRFs~\cite{tez1,Joz00}.

In addition to these pragmatic issues, the presumed necessity of SRFs
also touches upon a number of more foundational questions in quantum
mechanics. For instance, it has been argued~\cite{dickson} that the
physical nature of SRFs is the key issue in Bohr's reply to Einstein,
Podolsky and Rosen.  In this context, it is interesting that all
currently proposed schemes for violating Bell inequalities presume the
existence of a SRF -- the results presented here indicate that this
presumption is in fact unnecessary.

In this letter, we show that both classical and quantum communication
can be achieved \emph{without} first establishing a SRF by using
entangled states of multiple qubits. We explicitly describe a scheme
that employs two qubits to communicate a single classical bit of
information, and a scheme that uses four qubits to transmit an encoded
``logical'' qubit. In both schemes, the communication is achieved with
perfect fidelity. (Note that, in contrast, any scheme that attempts to
establish a SRF first, using a finite amount of communication, will
necessarily be subject to errors.)  We present the optimal schemes for
communicating classical and quantum information with perfect fidelity
given $N$ transmitted qubits, and we prove that communication of one
classical bit per transmitted qubit or one logical qubit per
transmitted qubit can be achieved asymptotically.  As an explicit
example of the practicality of our scheme for classical communication
without a SRF, we propose a feasible experiment using existing optical
technology to communicate one classical bit of information per
entangled photon pair.

Our communication scenario consists of two parties that have access to
a quantum channel but do not possess a SRF.  For simplicity, we
consider a noiseless channel that transmits qubits (our results can be
extended to noisy channels or higher-dimensional systems). Such a
channel defines an isomorphism between Alice's and Bob's local
experimental operations.  Specifically, representing Alice's
experimental operations using one qubit Hilbert space and Bob's using
another, the isomorphism is given by a unitary map $R(\Omega)$,
$\Omega \in$ SU(2) between them.  We define the lack of a SRF as a
lack of any knowledge of this isomorphism; i.e., a lack of any
knowledge of $\Omega$.  If Alice prepares a qubit in the state $\rho$
and transmits it to Bob, he represents the state of this received
qubit as a mixed density operator
\begin{equation}
  \label{eq:SingleQubitSuperop}
  \mathcal{E}_1(\rho) = \int {\rm d}\Omega\, R(\Omega) \rho R^\dag
  (\Omega) = \tfrac{1}{2}I \, ,
\end{equation}
obtained by averaging over all possible isomorphisms, i.e., all
unitary maps $R(\Omega)$, $\Omega \in$ SU(2).  (Here, ${\rm d}\Omega$
is the SU(2)-invariant measure.)  Thus, without a SRF, Alice cannot
communicate any information to Bob using only a single
qubit.

However, if Alice chooses to send more than one qubit to Bob, some
information \emph{can} be transmitted because the relative state of
the qubits carries information regardless of the existence of a SRF.
For instance, if Alice prepares two qubits in the state $\rho$, Bob
describes this same pair of qubits by the state that results by
application of the superoperator
\begin{equation}
  \label{eq:TwoQubitDecoheringChannel}
  \mathcal{E}_2(\rho) = \int {\rm d}\Omega \, R_1(\Omega)\otimes
  R_2(\Omega)  \rho R_1^\dag (\Omega)\otimes R_2^\dag(\Omega) \, .
\end{equation}
Note that this two-qubit superoperator does not average over
independent transformations for each qubit; instead, it averages over
a single qubit transformation $\Omega \in$ SU(2) applied identically
to both qubits.

Consider the following example where Alice encodes a single classical
bit $b$ by transmiting two qubits: for $b=0$, Alice sends parallel
spins ($|0\rangle_1 |0\rangle_2$), and for $b=1$ she sends
anti-parallel spins ($|0\rangle_1 |1\rangle_2$).  Using his optimal
measurement~\cite{Hel76}, Bob can correctly estimate $b$ with
probability $3/4$.  Thus, with this scheme, \emph{some} information
about Alice's bit is transmitted without a SRF, but some is lost.

However, Alice need not send product states as in the above example.
As with the problem of establishing a shared direction~\cite{Per01} or
Cartesian frame~\cite{Per01b}, entanglement between qubits provides an
advantage.  To determine which (possibly entangled) states may allow
for optimal communication, we note that the tensor representation of
SU(2) on two qubits decomposes into a direct sum of a $j=0$
irreducible representation (irrep) carried by the antisymmetric state
$|\Psi^-\rangle = \frac{1}{\sqrt{2}}(|01\rangle_{12} -
|10\rangle_{12})$ and a $j=1$ irrep carried by the symmetric states,
and that the tensor representation of SU(2) on this direct sum does
not mix these irreps.  Thus, the antisymmetric state is invariant
under the action of $\mathcal{E}_2$, $\mathcal{E}_2(|\Psi^-\rangle
\langle \Psi^-|) = |\Psi^-\rangle \langle \Psi^-|$, and any density
operator with support on the symmetric subspace is mapped by
$\mathcal{E}_2$ to the completely mixed state
$\frac{1}{3}\mathbb{I}_{j=1}$ over the symmetric subspace.  Thus, we
propose the following communication protocol.  Alice sends Bob the
antisymmetric state $|\Psi^-\rangle$ to communicate $b=0$ and
\emph{any} state in the symmetric subspace for $b=1$.  Bob then
performs a projective measurement onto the antisymmetric and symmetric
subspaces and will recover $b$ with certainty.  Thus, using this
protocol, Alice can communicate one classical bit to Bob for every two
qubits sent.

The efficiency of the scheme can be increased by entangling more
qubits.  Consider the transmission of $N$ qubits; the superoperator
$\mathcal{E}_N$ that describes the lack of a SRF acting on a general
density operator $\rho$ of $N$ qubits is given by
\begin{equation}
  \label{eq:NQubitDecoheringChannel}
  \mathcal{E}_N(\rho) = \int {\rm d}\Omega \, R_1(\Omega)\cdots
  R_N(\Omega)  \rho R_1^\dag (\Omega)\cdots R_N^\dag(\Omega) \, .
\end{equation}
This ``collective'' tensor representation of SU(2) on $N$ $j=1/2$
systems (i.e., $R(\Omega) \in$ SU(2) acting identically on all qubits)
can again be decomposed into a direct sum of SU(2) irreps, with
angular momentum quantum number $j$ ranging from $0$ or $1/2$ to
$N/2$.  In general, there will be multiple irreps for a given value of
$j$.  For simplicity, we assume that $N$ is even.  In this case, we
can express the resulting direct sum as
\begin{multline}
  \label{eq:DirectSum}
  {\rm SU}(2)_{1/2}^{\otimes N} = c^{(N)}_{N/2} {\rm SU}(2)_{N/2} \, \oplus \,
  c^{(N)}_{N/2-1} {\rm SU}(2)_{N/2-1} \, \oplus \cdots \\ \cdots \oplus\,
  c^{(N)}_0  {\rm SU}(2)_{0} \, ,
\end{multline}
where ${\rm SU(2)}_j$ denotes the irrep of SU(2) with angular momentum
quantum number $j$, and $c^{(N)}_j$ denotes the number of times that
the irrep ${\rm SU(2)}_j$ appears in the direct sum (i.e., the
multiplicity of the irrep).

We note that the different irreps of the same $j$ value (the
multiplicities) are defined by the ordering of the coupling, because
there are in general many ways to couple $N$ particles to total $j$.
Thus, to agree on the definitions of the multiple irreps for a given
$j$, Alice and Bob must agree on a choice of ordering of the coupling.
This agreement on coupling does not require a SRF, but does require
that Alice and Bob agree on a labeling $i \in (1,\ldots,N)$ of each
qubit.

We can now state and prove the result for classical communication.

\textbf{Proposition:} The maximum number of classical messages
that can be perfectly transmitted without a SRF is equal to the
number $C^{(N)}$ of SU(2) irreps in the direct sum decomposition
of the tensor representation of SU(2) on $N$ qubits.

\textbf{Proof:} We employ the following property of $\mathcal{E}_N$:
for any state $|\psi_{j,r}\rangle$ in the carrier space
$\mathbb{H}_{j,r}$ of the irrep labelled by $j,r$ (where $r$ is a
label for the multiplicity), the state $\rho_{j,r} \equiv
\mathcal{E}_N(|\psi_{j,r}\rangle \langle\psi_{j,r}|) = \frac{1}{2j+1}
\mathbb{I}_{j,r}$ is the completely mixed state over that irrep.  To
transmit $C^{(N)}$ classical messages, it is sufficient for Alice to
encode these messages using $C^{(N)}$ distinct states, one chosen from
each irrep.  Bob can perform a measurement associated with the
projector-valued measure $\{ \mathbb{I}_{j,r} \}$ to distinguish the
subspaces corresponding to the direct sum decomposition.  For Alice to
send an additional message, she must be able to prepare a state
$|\psi'\rangle$ that Bob can distinguish from the other states with
certainty.  Thus, $\rho' \equiv \mathcal{E}_N(|\psi'\rangle\langle
\psi'|)$ must be orthogonal to $\rho_{j,r}$ for all $j,r$.  There does
not exist such a $\rho'$ because the supports of the $\rho_{j,r}$ span
the entire Hilbert space.  \hfill$\Box$\medskip

To determine $C^{(N)}$, we note that the multiplicity $c^{(N)}_j$ of
each irrep in the direct sum decomposition is determined by the
dimension of the corresponding representation of the symmetric group
(the group of permutations of the $N$ systems)~\cite{Ful91}.  Thus,
$c^{(N)}_j$ can be calculated using Young tableaux: it is the number
of possible Young tableaux for a Young diagram consisting of two rows,
the first row consisting of $N/2 + j$ columns and the second
consisting of $N/2-j$ columns.  Using the hook lengths to calculate
the number of Young tableaux yields
\begin{equation}
  \label{eq:Multiplicity}
  c^{(N)}_j = \frac{N!}{\prod{\rm hook\ lengths}} =
  \binom{N}{N/2-j}\frac{2j+1}{N/2+j+1} \, .
\end{equation}
The total number of SU(2) irreps that appear in the direct sum
decomposition for $N$ qubits is
\begin{equation}
  \label{eq:TotalMult}
  C^{(N)} = \sum_{j=0}^{N/2} c^{(N)}_j = \binom{N}{N/2} \, .
\end{equation}
The number of classical bits that can be transmitted per qubit using
the above scheme is $N^{-1} \log_2 C^{(N)}$, which tends
asymptotically to $1-(2N)^{-1}\log_2 N$.  Thus, in the large $N$
limit, one classical bit can be transmitted for every qubit sent.
Remarkably, this rate is equivalent to what can be accomplished if
Alice and Bob \emph{do} possess a SRF.

In general, the states to be transmitted in the optimal scheme for $N$
qubits are highly entangled.  (They include, for example, singlet
states of $N$ qubits.)  Such multipartite entangled states are
difficult to prepare in practice. However, as we now show, for the
case $N=2$ the required entanglement is easily achieved using quantum
optics, in particular using the polarization degree of freedom of a
photon.

When using an optical fibre for transmitting polarized photons, Bob
typically has no knowledge of the relationship between Alice's
polarization axes and his own.  Such an optical fibre is an instance
of a quantum channel without a SRF.  To demonstrate communication of a
single classical bit using such a channel, we can make use of
maximally-entangled photon pairs (Bell states) produced using
parametric downconversion (PDC)~\cite{Kwi95}.  For example, by
selecting two spatial modes (each with two polarization states,
$|H\rangle$ and $|V\rangle$) from the PDC output, one can prepare the
antisymmetric Bell state $|\Psi^-\rangle_{12} = \frac{1}{\sqrt{2}}
(|H\rangle_1|V\rangle_2 - |V\rangle_1|H\rangle_2)$.  By performing a
$90^\circ$ polarization rotation on one of the spatial modes, Alice
can also prepare the symmetric Bell state $|\Phi^-\rangle_{12} =
\frac{1}{\sqrt{2}} (|H\rangle_1|H\rangle_2 - |V\rangle_1|V\rangle_2)$.
Thus, in our proposed experiment, Alice prepares the antisymmetric
state $|\Psi^-\rangle_{12}$ to encode the classical bit $b=0$, and
prepares $|\Phi^-\rangle_{12}$ for $b=1$.

Alice then transmits these photons to Bob, who performs a projective
measurement onto the antisymmetric and symmetric subspaces of the two
spatial modes in order to retrieve the classical bit $b$. To perform
this measurement, Bob employs a linear optics Bell state analyzer. An
ideal Bell state analyzer that distinguishes all four Bell states is
impossible using only linear optics and photodetectors~\cite{Lut99};
however, such a complete measurement is not required in this example.
Using linear optics, one \emph{can} distinguish the antisymmetric
state from the symmetric ones~\cite{Wei94}.  Such a measurement scheme
employs a 50/50 beam splitter that mixes the two spatial modes,
followed by photodetection at each output mode.  A coincidence
detection (each photodetector detects one photon) indicates the
antisymmetric state, whereas the detection of two photons at a single
photodetector indicates a symmetric state.  Thus, using existing
quantum optics technology, it is possible to communicate classical
bits using entangled photon pairs without a SRF.

We now turn to the problem of \emph{quantum} communication in the
absence of a SRF.  It is clear from Eq.~(\ref{eq:SingleQubitSuperop})
that a single transmitted qubit can convey no quantum information.
However, in analogy to our classical communication results, multiple
transmitted qubits do allow for this possibility.  Although quantum
information can only be communicated with imperfect fidelity using two
transmitted qubits, we now demonstrate that perfect fidelity can be
achieved by using more than two qubits.

The key insight is that, because $\mathcal{E}_N$ describes a
collective decoherence mechanism, we can appeal to the techniques of
decoherence free subspaces (DFSs)~\cite{Zan97}.  For $N$ (even)
transmitted qubits, we observe that the superoperator $\mathcal{E}_N$
leaves all $j=0$ states in the direct sum decomposition invariant.
Thus, the $j=0$ states span a DFS, denoted $\mathbb{H}_{\rm DFS}$.
The number of $j=0$ states is given by the multiplicity ${\rm dim}\ 
\mathbb{H}_{\rm DFS} = c^{(N)}_0 = \binom{N}{N/2} \frac{1}{N/2+1}$.

For $N=2$, there is only one $j=0$ state: the Bell state
$|\Psi^-\rangle$.  Since no quantum information can be encoded in a
one-dimensional subspace, two physical qubits are insufficient for the
purpose of transmitting quantum information with perfect fidelity.
For $N=4$, on the other hand, there are two distinct $j=0$ states,
specifically,
\begin{align}
  \label{eq:TwoSinglets}
  |0_L\rangle &= \tfrac{1}{2}(|01\rangle_{12} -
   |10\rangle_{12}) (|01\rangle_{34} -
   |10\rangle_{34}) \\
  |1_L\rangle &= \tfrac{1}{\sqrt{3}}(|00 1
  1\rangle_{1234}  + |11 0
   0\rangle_{1234}) \nonumber \\
  &\quad - \tfrac{1}{2\sqrt{3}}(|01\rangle_{12} +
   |10\rangle_{12}) (|01\rangle_{34} +
   |10\rangle_{34}) \, , 
\end{align}
where $\{|0\rangle,|1\rangle\}$ is \emph{any} orthogonal basis for the
single qubit Hilbert space.  The superoperator $\mathcal{E}_N$
preserves the two-dimensional subspace spanned by these states, i.e.,
this subspace is a DFS.

Thus four physical qubits can encode a single logical qubit.
Single-qubit operations on this logical qubit are an encoded
representation of SU(2) that commutes with the superoperator
$\mathcal{E}_N$.  The encoded generators are given by Hermitian
exchange operations (i.e., two-qubit permutations), which clearly do
not require a SRF; for details of the encoded SU(2) group as well as
two-logical-qubit coupling operations, see~\cite{Zan97,Kni00}.

\emph{Noiseless subsystems}~\cite{Kni00} can be used to maximize the
amount of encoded quantum information protected from the decohering
superoperator $\mathcal{E}_N$.  For example, it is possible to encode
one logical qubit into only three physical qubits.  For a given number
$N$ of physical qubits, the maximal subsystem is given by the irrep
$j_{\rm max}$ with the greatest multiplicity $c^{(N)}_j$.
Asymptotically, this irrep is found to be $j_{\rm max} = \sqrt{N}/2$,
and the number $N^{-1} \log_2 c^{(N)}_{j_{\rm max}}$ of logical qubits
encoded per physical qubit in $N$ physical qubits behaves as $1 -
N^{-1}\log_2 N$, approaching unity for large $N$.  This remarkable
result proves that quantum communication without a SRF is
asymptotically as efficient as quantum communication with a SRF, and
is the communication analog of ``asymptotic
universality''~\cite{Kni00}.

These results imply that Alice and Bob can share entangled states in
the absence of a reference frame.  For instance, Alice can prepare two
quadruplets of physical qubits in the state
$\frac{1}{\sqrt{2}}(|0_L0_L\rangle+|1_L1_L\rangle)$, and send one
quadruplet to Bob. Since Alice and Bob can perform any measurement in
their respective logical qubit Hilbert spaces, they can violate Bell
inequalities despite having no SRF.  It also follows that such
entangled states can be used for quantum teleportation, which implies
that the latter does not rely upon the existence of a SRF either,
contrary to the claims of~\cite{Enk01}.

Another situation of interest is if Alice and Bob share a
\emph{partial} reference frame, for instance, if they share only a
single direction in space rather than a full Cartesian frame.  In this
case, the superoperator describing a partial SRF corresponds to what
the DFS community calls a collective dephasing operation~\cite{Dua98}.
Here, Alice and Bob can obviously transmit a classical bit using a
single qubit.  To communicate a single logical qubit, it suffices to
transmit two physical qubits, and asymptotically, the ratio of logical
qubits to transmitted qubits is $1-(2N)^{-1}\log_2 N$.

We note that the encoding used in our schemes also protects against
channel noise that affects all qubits identically~\cite{Zan97}. If all
transmitted qubits are sent close together in space and time, such a
description will be appropriate.  It follows, in particular, that
noise in the evolution of transmitted qubits, which is problematic for
the quantum clock synchronization protocol of~\cite{Joz00}, will
generally not cause errors in our communication schemes.  On the other
hand, a noisy channel that affects the individual transmitted qubits
differently or that causes a loss of information about the ordering of
the qubits \emph{will} be problematic.  However, concatenated
encodings and quantum error correction can be used to accommodate this
noise.

There remain many interesting questions about the role of reference
frames in quantum theory. For instance, it appears that the
availability of a reference frame for some degree of freedom
determines whether or not it is appropriate to assume a superselection
rule for the complementary variable (the status of such rules has been
the subject of some controversy~\cite{Aha67}).  Another problem of
interest is to determine how these results generalize to relativistic
quantum mechanics, wherein reference frames have particular
significance.

In conclusion, we have shown how to perform both classical and quantum
communication without a SRF, thereby proving that a SRF is not a
necessary requirement for communication or distributed quantum
information processing.  Also, we have shown that asymptotically this
communication can be performed as efficiently as if a SRF was
available.  We have proposed an experiment to demonstrate this
principle using entangled photons.

\begin{acknowledgments}
  This project has been supported by the Australian Research Council
  and Macquarie University.  T.R.\ is supported by the NSA \& ARO
  under contract No.\ DAAG55-98-C-0040.  R.W.S. is supported in part
  by NSERC of Canada.  We acknowledge helpful discussions with D.\ 
  Berry, R.\ Mu\"noz-Tapia, M.\ A.\ Nielsen, D.\ J.\ Rowe, B.\ C.\ 
  Sanders, S.\ van Enk, F.\ Verstraete and H.\ M.\ Wiseman.
\end{acknowledgments}

\end{document}